\documentclass[11pt]{article}
\usepackage{moriond}
\usepackage{graphicx}
\usepackage{epstopdf}

\def\Journal#1#2#3#4{{#1} {\bf #2}, #3 (#4)}

\def\be{\begin{equation}}
\def\ee{\end{equation}}
\def\bea{\begin{eqnarray}}
\def\eea{\end{eqnarray}}

\newcommand{\Photo}{\includegraphics[height=35mm]{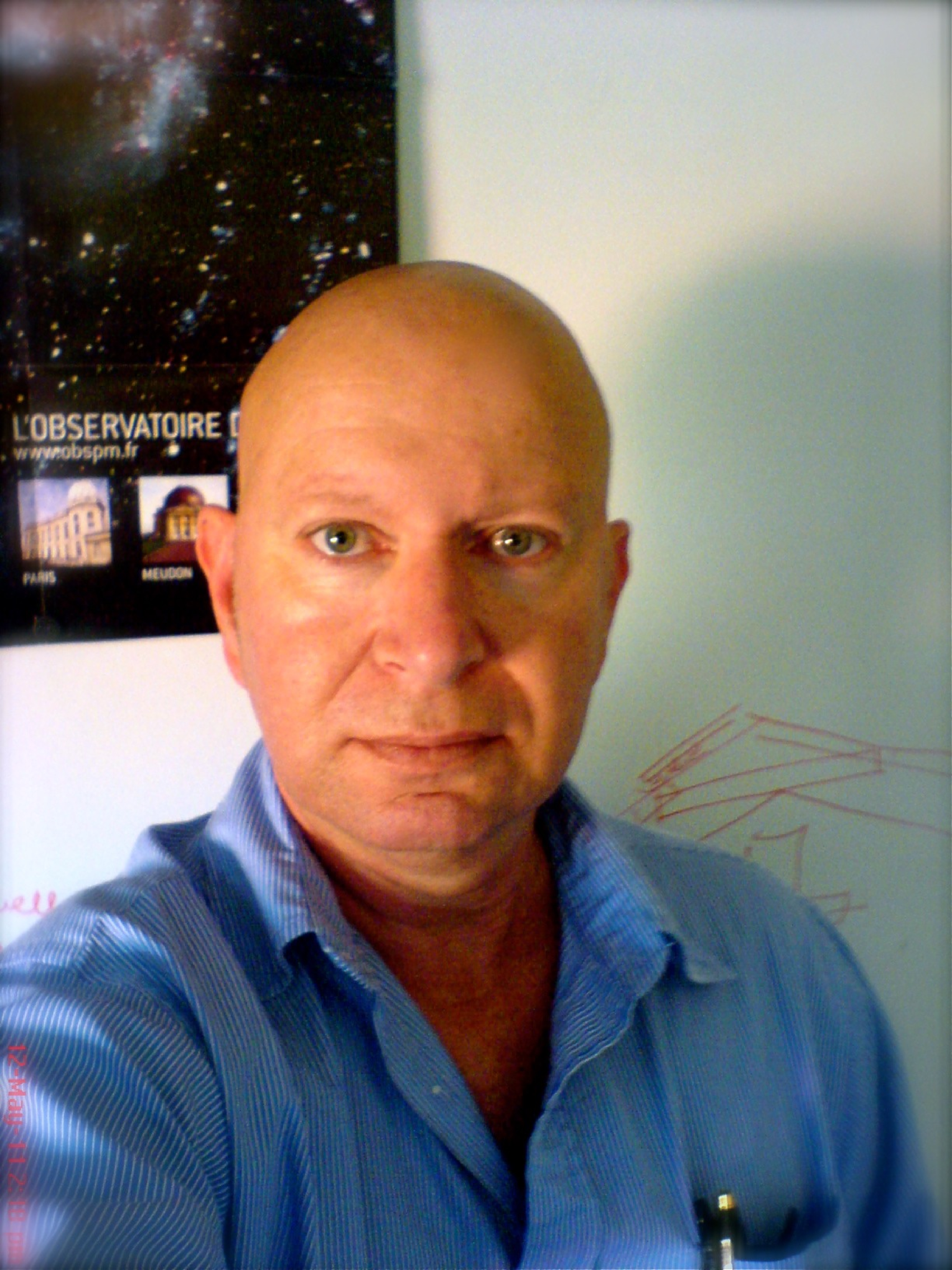}}

\begin{document}
\vspace*{4cm}
\title{The Swift Gamma-Ray Burst redshift distribution: selection biases or rate evolution at high-z?}

\author{ D.M. Coward$^{1}$, E.J. Howell$^{1}$,  M. Branchesi$^{2}$, B. Gendre$^{3}$, G. Stratta$^{3}$}
\address{$^{1}$School of Physics, University of Western Australia, Crawley WA 6009, Australia\\
$^{2}$DiSBeF - Universit\`a degli Studi di Urbino `Carlo Bo', I-61029 Urbino, Italy\\
$^{3}$INAF - Osservatorio Astronomico di Roma, Via Frascati 33, I-00040 Monteporzio Catone (Roma), Italy}

\maketitle\abstracts{
We employ realistic constraints on selection effects to model the Gamma-Ray Burst (GRB) redshift distribution using {\it Swift} triggered redshift samples acquired from optical afterglows and the TOUGH survey. Models for the Malmquist bias, redshift desert, and the fraction of afterglows missing because of host galaxy dust extinction, are used to show how the ``true'' GRB redshift distribution is distorted to its presently observed biased distribution. Our analysis, which accounts for the missing fraction of redshifts in the two data subsets, shows that a combination of selection effects (both instrumental and astrophysical) can describe the observed GRB redshift distribution. The observed distribution supports the case for dust extinction as the dominant astrophysical selection effect that shapes the redshift distribution. }

\section{Introduction}
In this study, we use realistic constraints and models for redshift dependant selection biases, combined with GRB OA luminosities, to show how selection effects distort the ``trure'' spatial distribution to its presently observed distribution. We employ two subsets of GRB redshifts. The first, Howell \& Coward (2013) \cite{2013MNRAS.428..167H}, hereafter HC, uses 141 {\it Swift} triggered spectroscopic absorption redshifts from OAs up to Oct 2012. \footnote{GRB redshift sample is a subset taken from GCN circulars and http://www.mpe.mpg.de/$\sim$jcg/grbgen.html}. The second, less biased but smaller sample, uses a subset of 58 redshifts from the TOUGH (The Optically Unbiased GRB Host) survey \cite{2012ApJ...756..187H}. By accounting for selection effects, we investigate if the observed GRB redshift distribution is compatible with GRB rate evolution tracking the global star formation rate.

\subsection{GRB optical selection effects}
We define gamma ray burst (GRB) optical afterglow (OA) selection effects as the combination of sensitivity limited optical follow-up and phenomena, astrophysical and instrumental, that reduce the detection probability of an OA. Some of the more widely understood effects are discussed by Fynbo et al.\cite{2009ApJS..185..526F} in detail. Optical biases have reduced the fraction of {\it Swift} triggered OAs, and have introduced a selection towards detecting the brightest OAs, hence the more nearby bursts. Additionally, there are biases that distort the redshift distribution over certain redshift ranges (see e.g. \cite{cow08,cow09}). See Coward et al. \cite{cow13} for a complete description:\\
\begin{enumerate}

\item {\bf Malmquist bias:} This bias arises because the telescopes and instruments acquiring OA absorption spectra (and photometry) are limited by sensitivity. In reality, the instruments acquiring redshifts are biased to sampling the bright end of the OA luminosity function. To account for this bias, it is necessary to have some knowledge of OA luminosity function (which is uncertain especially at the faint end), and an estimate of the average sensitivity limit of the instruments. This is the most fundamental bias that encompasses all flux limited detection and is the basis for modelling a selection function for OA/redshift measurement.


Kann et al. (2010) \cite{2010ApJ...720.1513K} find a weak correlation between $E_{{\rm iso}}$ and $L_{{\rm opt}}$ with a Kendall rank correlation coefficient of 0.29 for the subset of GRBs used in this study to estimate an optical LF. The key question is: how does this correlation affect the Malmquist bias? Firstly consider the effect if $E_{{\rm iso}}$ and $L_{{\rm opt}}$ are uncorrelated. This implies that a high $E_{{\rm iso}}$ (that will be preferentially detected by {\it Swift}) could be associated with equal probability with either a low or high luminosity optical afterglow. This scenario implies that detection of the OA is independent of the high energy luminosity. Alternatively, observation suggests a high $E_{{\rm iso}}$ preferentially selects a high optical luminosity. This implies that the Malmquist bias will be reduced at high-$z$, because it is the high $E_{{\rm iso}}$ bursts that are preferentially seen at large-$z$. These bursts will also be more optically luminous, so that redshift measurement will be more probable.

\item {\bf Redshift desert:} The so-called redshift desert is a region in redshift ($1.4 < z < 2.5$) where it is difficult to measure absorption and emission spectra. 
As redshift increases beyond $z\sim 1$, the main spectral features become harder to recognize as they enter a wavelength region where the sensitivity of CCDs starts to drop and sky brightness increases. Beyond $z\sim1.4$, the spectral features move beyond 1 $\mu$m, i.e., into the near-IR.  In the case of actively starforming galaxies at $z> 1.4$, these are several narrow absorption lines over the UV continuum, most of which originate in the ISM of these galaxies. 

\item {\bf Different redshift measurement techniques:} Historically, because of the deficiency in pre-{\it Swift} ground-based follow-up of GRBs, there was a strong bias for imaging the brightest bursts. Because the brightest bursts are predominantly nearby, a significant fraction of the first GRB redshifts were obtained by emission spectroscopy of the host galaxy.
In the {\it Swift} era (from 2005 onwards), an optical afterglow (OA) is usually required to measure a redshift. For most high-$z$ GRBs, this is achieved by absorption spectroscopy of the GRB afterglow. The host galaxies are usually too faint to make a significant contribution to the spectra. Most GRB spectroscopic redshifts are acquired by large aperture ground based telescopes, including VLT, Gemini-S-N, Keck and Lick (see \cite{2009ApJS..185..526F} for a more complete list along with specific spectroscopy instruments). The measurement of a GRB redshift depends strongly on the limiting sensitivity and spectral coverage of the spectroscopic system. This bias is expected to manifest at high-$z$, where the optically brightest OAs are near the limiting sensitivity of the telescope. 

\item {\bf Host galaxy extinction:} 
 There has been growing evidence that dark bursts are obscured in their host galaxies e.g. \cite{jakob04,2009AJ....138.1690P}. These studies generally show that GRBs originating in very red host galaxies always show some evidence of dust extinction in their afterglows. Also a significant fraction of dark burst hosts have extinction columns with $A_V \sim1$ mag, and some as high as $A_V =2-6$ mag  \cite{2009AJ....138.1690P}. \\
 
 \begin{figure}
\centering
\includegraphics[scale=0.5]{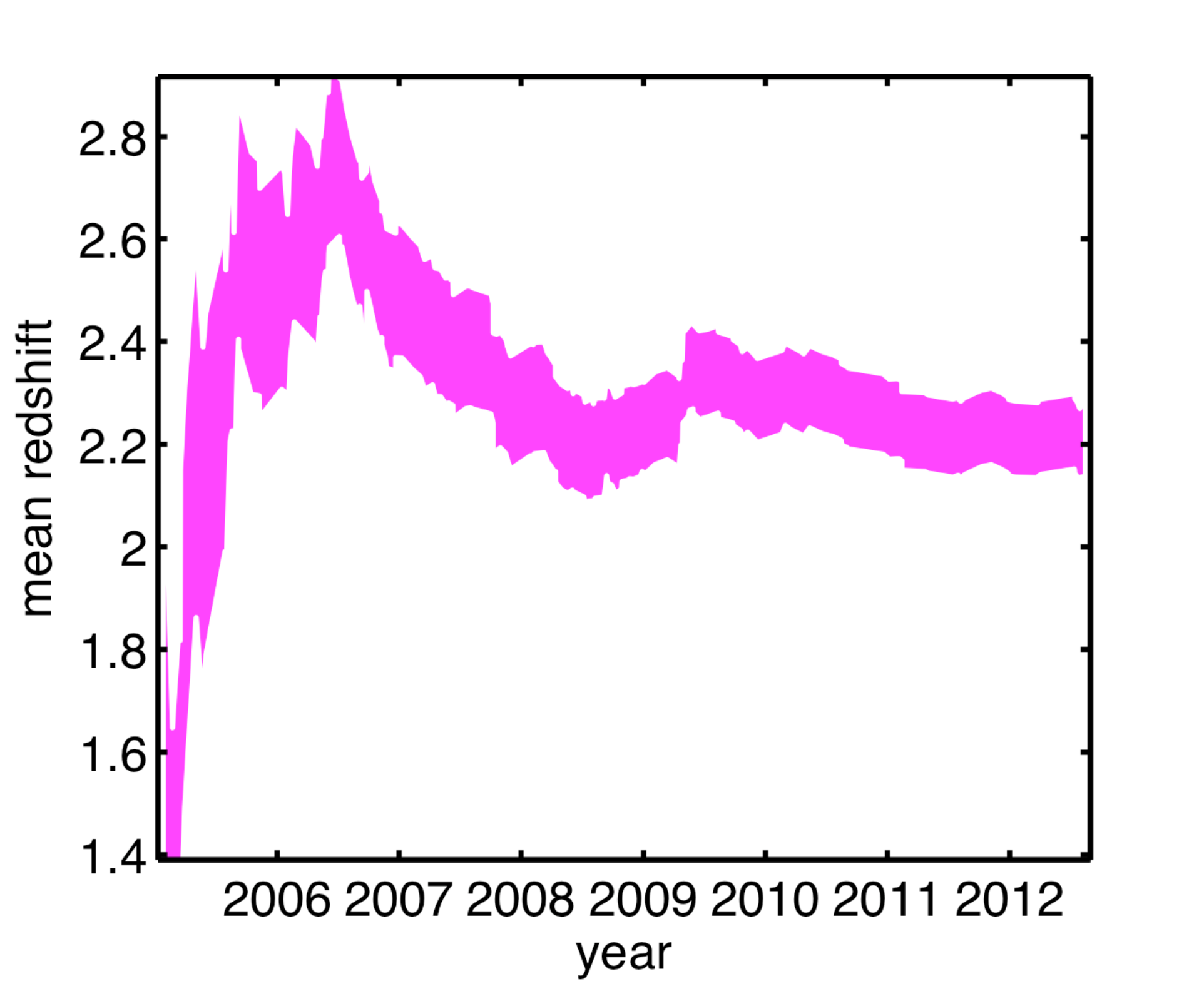}
\caption{The {\it Swift} mean redshift uncertainty bound plotted over the duration of the mission. It is clear there is a drift in the mean redshift over time, a consequence of different priorities and instruments contributing to redshift acquisition. The {\it jump} observed in 2009 is a result of GRBs 090423 and 090429B, with redshifts of $z=8.26$ (NIR spectroscopic) and $z=9.2$ (photometric) respectively. } \label{zmean}
\end{figure}

\end{enumerate}

\section{GRB redshift distribution model with selection effects}\label{prob}
 
The dominant GRB redshift distribution biases discussed above are represented as the product of independent dimensionless selection functions that are unity for a 100\% selection probability (see Coward et al. \cite{cow13}): 
\begin{enumerate}
\item $\psi_{\rm Obs}$ -- number dropouts from mostly non-redshift dependant biases, which are different depending on the selection criteria for the sample. We assume that the TOUGH sample is relatively free of instrumental biases, but about $20\%$ of the HC sample is affected by instrumental biases.     
\item $\psi_{\rm Swift}(z)$ -- the limited sensitivity of {\it Swift} to trigger on GRBs.
\item  $\psi_{\rm M}(z)$ -- the limited sensitivity of instruments to measure a redshift from the GRB OA.
\item $\psi_{\rm Desert}(z)$ -- number dropouts from the redshift desert.
\item $\psi_{\rm Dust}(z)$ -- number dropouts from host galaxy dust extinction. 
\end{enumerate}

The GRB redshift probability distribution function, that includes the above selection effects, can be expressed as: 
\begin{equation}
\begin{array}{ll}
P(z) = & N_p\frac{dV(z)}{dz}\frac{e(z)}{(1+z)} \psi_{\rm Swift}(z)\psi_{\rm Obs} \psi_{\rm M}(z)\\
&  \psi_{\rm Desert}(z) \psi_{\rm Dust}(z)
\end{array}
\end{equation}
where $N$ is a normalization constant. The volume element, $dV/dz$, is calculated using a flat-$\Lambda$ cosmology with $H_{\mathrm 0}$ = 71 km s$^{-1}$ Mpc$^{-1}$, $\Omega_M$ = 0.3 and $\Omega_\Lambda$ = 0.7, and we fix $\psi_{\rm Obs}\approx0.5$ (see the selection effects listed above). The function $e(z)$ is the dimensionless source rate density evolution function (scaled so that $e(0)=1$). We assume that $e(z)$ tracks the star formation rate history (See Coward et al. \cite{cow13} for a full description of the model):

\begin{figure}

\includegraphics[scale=0.7]{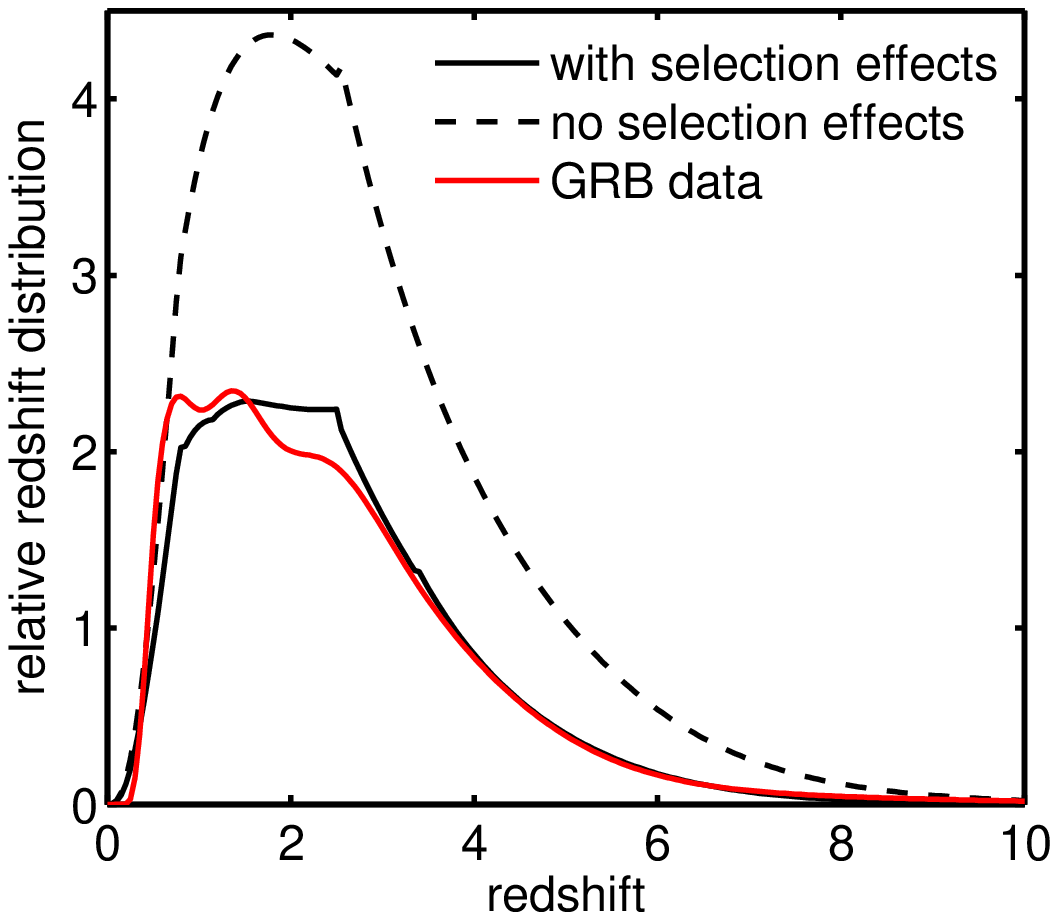}
\includegraphics[scale=0.7]{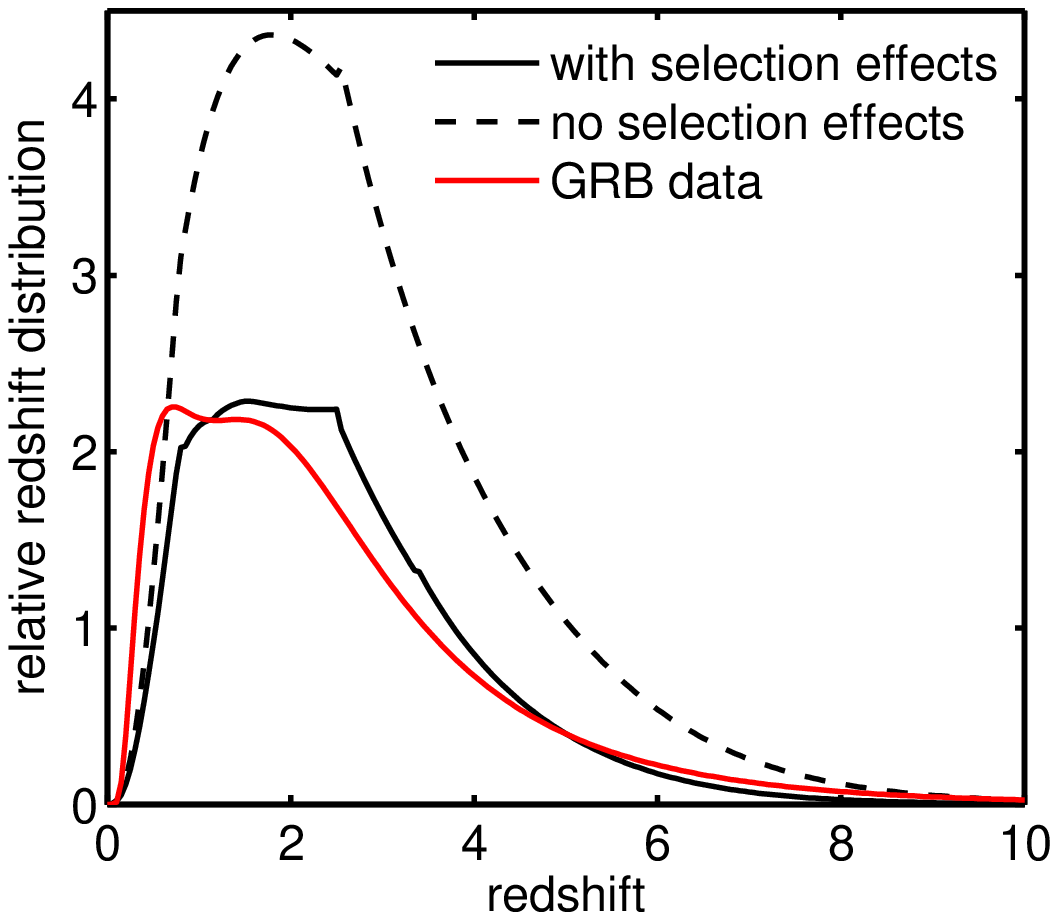}
\caption{{\bf Left:} The redshift distributions for models with and without optical selection effects, and the relative distribution of the HC sample. The optimal models, defined as having both high K-S probabilities and small fractional errors, are Models which includes selection effects, a GRB rate evolution at $z=10$ similar to that of $z=1$, and either including or excluding a Malmquist bias correction. The least optimal model excludes selection effects. {\bf Right:} Same as the left figure but using the TOUGH redshift distribution. Both the HC and TOUGH data require the same optimal models that include selection effects.} \label{results}
\end{figure}

\section{Summary}
Fig. \ref{results} plots the observed redshift distribution, with the optimal model (that includes selection effects), and for comparison the expected distribution that would be observed if all optical selection effects were removed.
In summary, our analysis suggests that a combination of selection effects (both instrumental and astrophysical) can adequately describe the observed redshift distribution. Furthermore the observed distribution is compatible with a rate evolution that tracks the evolving SFR. We show that the TOUGH selection and a subset of absorption redshifts (the HC sub-sample) are compatible and both support the case for dust extinction as the dominant astrophysical selection effect that shapes the redshift distribution.

\section*{Acknowledgments}
D.M. Coward is supported by an Australian Research Council Future Fellowship. E.J. Howell acknowledges support from a University of Western Australia Fellowship.

\section*{References}


\end{document}